\documentclass[runningheads]{llncs}

\newif\ifAnonymous
\newif\ifToolDetails
\Anonymousfalse
\ToolDetailstrue

\usepackage[hyphens]{url}
\usepackage[breaklinks,hidelinks]{hyperref}
\usepackage[anythingbreaks]{breakurl}

\usepackage{verbatim}
\usepackage{multirow}

\usepackage[normalem]{ulem}

\usepackage{tikz}
\usepackage{listings,color}
\usetikzlibrary{shadows,shapes,shapes.geometric,arrows}
\usepackage{pgfplots}
\pgfplotsset{compat=1.10}

\definecolor{dkgreen}{rgb}{0,0.6,0}
\definecolor{forestgreen}{RGB}{0,100,50}
\definecolor{redd}{RGB}{76,0,153}
\definecolor{english}{rgb}{0.0, 0.26, 0.15}
\definecolor{coolblack}{rgb}{0.0, 0.18, 0.39}
\definecolor{blue-violet}{rgb}{0.54, 0.17, 0.89}

\lstdefinestyle{customjava}
{language=java,
keywordstyle=\color{blue-violet},
basicstyle=\ttfamily\scriptsize,
morekeywords={String, var, require, msg, keccak256, now, function, continue, else},
mathescape=true,
escapeinside={/*@}{@*/},
commentstyle=\color{english},   
keywordstyle=[2]{\color{red}},
}

\lstdefinestyle{customjavaexample}
{language=java,
keywordstyle=\color{blue-violet},
basicstyle=\ttfamily\scriptsize,
morekeywords={String, var, require, msg, keccak256, now, function, continue, else},
mathescape=true,
escapeinside={/*@}{@*/},
commentstyle=\color{english},   
keywordstyle=[2]{\color{red}},
frame=single
}

\usepackage[textsize=tiny,textwidth=4cm,disable]{todonotes}
\usepackage{amsmath}
\usepackage{amsfonts}
\usepackage{paralist}

\newcommand{\Natassa}[1]{\todo[color=red!10,linecolor=black!50]{\textbf{Natassa}: #1}}
\newcommand{\Aron}[1]{\todo[color=blue!5,linecolor=black!50]{\textbf{Aron}: #1}}

\begin{document}
\setlength{\marginparwidth}{4cm}

\title{Designing Secure Ethereum Smart Contracts:\\A Finite State Machine Based Approach}
\titlerunning{Designing Secure Ethereum Smart Contracts}

\ifAnonymous
  \author{\vspace{-3em}} 
  \authorrunning{Anonymous}
  \institute{}
\else
  \author{Anastasia Mavridou\inst{1} \and Aron Laszka\inst{2}}
  \authorrunning{Anastasia Mavridou and Aron Laszka}
  \institute{Vanderbilt University \and University of Houston}
\fi

\maketitle

\ifAnonymous
\else
  \begin{center}
  Accepted for publication in the proceedings of the 22nd International Conference on Financial Cryptography and Data Security (FC 2018).
  \end{center}
\fi

\begin{abstract}
The adoption of blockchain-based distributed computation platforms is growing fast.
Some of these platforms, such as Ethereum, provide support for implementing smart contracts, which are envisioned to have novel applications in a broad range of areas, including finance and Internet-of-Things.
However, a significant number of smart contracts deployed in practice suffer from security vulnerabilities, which enable malicious users to steal assets from a contract or to cause damage.
Vulnerabilities present a serious issue since contracts may handle financial assets of considerable value, and contract bugs are non-fixable by design.
To help developers create more secure smart contracts,
we introduce \emph{FSolidM}, a framework rooted in rigorous semantics for designing contracts as Finite State Machines (FSM).
We present a tool for creating FSM on an easy-to-use graphical interface and for automatically generating Ethereum contracts.
Further, we introduce a set of design patterns, which we implement as plugins that developers can easily add to their contracts to enhance security and functionality.
\end{abstract}


\section{Introduction}
\label{sec:intro}

The adoption and importance of blockchain based distributed ledgers are growing fast.
For example, the market capitalization of Bitcoin, the most-popular cryptocurrency, has exceeded \$70 billion in 2017.\footnote{\url{https://coinmarketcap.com/currencies/bitcoin/}}
While the first generation of blockchain systems were designed to provide only cryptocurrencies, later systems, such as Ethereum, can also function as distributed computing platforms~\cite{underwood2016blockchain,wood2014ethereum}.
These distributed and trustworthy  platforms enable the implementation smart contracts, which can automatically execute or enforce their contractual terms~\cite{clack2016smart}.
Beyond financial applications, blockchains are envisioned to have a wide range of applications, such as asset tracking for the Internet-of-Things~\cite{christidis2016blockchains}.
Due to their unique advantages,
blockchain based platforms and smart contracts are embraced by an increasing number of organizations and companies.
For instance, the project HyperLedger\footnote{\url{https://www.hyperledger.org/}}, which aims to develop open-source blockchain tools, is backed by major technology companies and financial firms, such as IBM, Cisco, J.P. Morgan, and Wells Fargo~\cite{vukolic2017rethinking}.

At the same time, smart contracts deployed in practice are riddled with bugs and security vulnerabilities.
A recent automated analysis of 19,336 smart contracts deployed on the public Ethereum blockchain found that 8,333 contracts suffer from at least one security issue~\cite{luu2016making}.
While not all of these issues lead to security vulnerabilities, many of them enable cyber-criminals to steal digital assets, such as cryptocurrencies.
For example, the perpetrator(s) of the infamous 2016 ``The DAO'' attack exploited a combination of vulnerabilities to steal 3.6 million Ethers, which was worth around \$50 million at the time of the attack~\cite{finley2016million}.
More recently, \$31 million worth of Ether was stolen due to a critical security flaw in a digital wallet contract~\cite{qureshi2017hacker}.
Furthermore, malicious attackers might be able to cause damage even without stealing any assets, e.g., by leading a smart contract into a deadlocked state, which does not allow the rightful owners to spend or withdraw their assets.

Security vulnerabilities in smart contracts present a serious issue for multiple reasons.
Firstly, smart contracts deployed in practice handle financial assets of significant value.
For example, at the time of writing, the combined value held by Ethereum contracts deployed on the public blockchain is 12,205,760 Ethers, which is worth more than \$3 billion.\footnote{\url{https://etherscan.io/accounts/c}}
Secondly, smart-contract bugs cannot be patched. 
By design, once a contract is deployed, its functionality cannot be altered even by its creator.
Finally, once a faulty or malicious transaction is recorded, it cannot be removed from the blockchain (``code is law'' principle~\cite{bhargavan2016short}).
The only way to roll back a transaction is by performing a hard fork of the blockchain, which requires consensus among the stakeholders and undermines the trustworthiness of the platform~\cite{leising2017ether}.

In practice, these vulnerabilities often arise due to the semantic gap between the assumptions contract writers make about the underlying execution semantics and the actual semantics of smart contracts~\cite{luu2016making}.
Prior work focused on addressing these issues in existing contracts by providing tools for verifying correctness~\cite{bhargavan2016short} and for identifying common vulnerabilities~\cite{luu2016making}.
%
%
In this paper, we explore a different avenue by proposing and implementing \emph{FSolidM}, a novel framework for creating secure smart contracts:
\begin{compactitem}
\item We introduce a formal, finite-state machine (FSM) based model for smart contracts. We designed our model primarily to support Ethereum smart contracts, but it may be applied on other platforms as well.
\item We provide an easy-to-use graphical editor that enables developers to design smart contracts as FSMs.
\item We provide a tool for translating FSMs into Solidity code.\footnote{Solidity is the most widely used high-level language for developing Ethereum contracts. Solidity code can be translated into Ethereum Virtual Machine bytecode, which can be deployed and executed on the platform.}
\item We provide a set of plugins that implement security features and design patterns, which developers can easily add to their model.
\end{compactitem}
Our tool is open-source and available online (see Section~\ref{sec:tool} for details).

The advantages of our approach, which aims to help developers create secure contracts rather than to fix existing ones, are threefold.
First, we provide a formal model with clear semantics and an easy-to-use graphical editor, thereby decreasing the semantic gap and eliminating the issues arising from it.
Second, rooting the whole process in rigorous semantics allows the connection of our framework to formal analysis tools~\cite{bensalem2009d,cavada2014nuxmv}.
Finally, our code generator---coupled with the plugins provided in our tool---enables developers to implement smart contracts with minimal amount of error-prone manual coding.

The remainder of this paper is organized as follows.
In Section~\ref{sec:related}, we give a brief overview of related work on smart contracts and common vulnerabilities.
In Section~\ref{sec:FSM}, we first present blind auction as a motivating example problem, which can be implemented as a smart contract, and then introduce our finite-state machine based contract model. In Section~\ref{sec:transformation}, we describe our FSM-to-Solidity code transformation.
In Section~\ref{sec:plugins}, we introduce plugins that extend the contract model with additional functionality and security features.
In Section~\ref{sec:tool}, we describe our FSolidM tool and provide numerical results on computational cost.
Finally, in Section~\ref{sec:concl}, we offer concluding remarks and outline future work.

\section{Related Work}
\label{sec:related}

\subsection{Common Vulnerabilities and Design Patterns}
\label{sec:commonVuln} 

Multiple studies investigate and provide taxonomies for common security vulnerabilities and design patterns in Ethereum smart contracts.
In Table~\ref{tab:common}, we list the vulnerabilities that we address and the patterns that we implement in our framework using plugins.

\begin{table} [t]
\setlength{\tabcolsep}{0.75em}
\renewcommand{\arraystretch}{1.2}
\caption{Common Smart-Contract Vulnerabilities and Design Patterns}
\label{tab:common}
\centering
\begin{tabular}{|l|l|l|}
\hline
Type & Common Name & FSolidM Plugin \\
\hline\hline
\multirow{3}{*}{Vulnerabilities} & reentrancy~\cite{luu2016making,atzei2017survey} & locking (Section~\ref{sec:locking}) \\ 
\cline{2-3}
& transaction ordering~\cite{luu2016making} & transition counter (Section~\ref{sec:transCounter}) \\
& a.k.a. unpredictable state~\cite{atzei2017survey} & \\
\hline
\multirow{2}{*}{Patterns} & time constraint~\cite{bartoletti2017empirical} & timed transitions (Section~\ref{sec:timedTrans}) \\
\cline{2-3}
& authorization~\cite{bartoletti2017empirical} & access control (Section~\ref{sec:accessContr}) \\
\hline
\end{tabular}
\end{table}

Atzei et al.\ provide a detailed taxonomy of security vulnerabilities in Ethereum smart contracts, identifying twelve distinct types~\cite{atzei2017survey}.
For nine vulnerability types, they show how an attacker could exploit the vulnerability to steal assets or to cause damage. 
Luu et al.\ discuss four of these vulnerability types in more detail, proposing various techniques for mitigating them (see Section~\ref{sec:verification})~\cite{luu2016making}.
In this paper, we focus on two types of these common vulnerabilities:
\begin{compactitem}
\item Reentrancy Vulnerability: Reentrancy is one of the most well-known vulnerabilities, which was also exploited in the infamous ``The DAO'' attack.
In Ethereum, when a contract calls a function in another contract, the caller has to wait for the call to finish. 
This allows the callee, who may be malicious, to take advantage of the intermediate state in which the caller is, e.g., by invoking a function in the caller.
\item Transaction-Ordering Dependence:
If multiple users invoke functions in the same contract, the order in which these calls are executed cannot be predicted.
Consequently, the users have uncertain knowledge of the state in which the contract will be when their individual calls are executed. 
\end{compactitem}

Bartoletti and Pompianu 
identify nine common design patterns in Ethereum smart contracts, 
and measure how many contracts use these patterns in practice~\cite{bartoletti2017empirical}.
Their results show that the two most common patterns are \emph{authorization} and \emph{time constraint}, which are used in 61\% and 33\% of all contracts, respectively. 
The also provide a taxonomy of Bitcoin and Ethereum contracts, dividing them into five categories based on their application domain.
Based on their categorization, they find that the most common Ethereum contracts deployed in practice are financial, notary, and games.

\subsection{Verification and Automated Vulnerability Discovery}
\label{sec:verification}

Multiple research efforts attempt to identify and fix these vulnerabilities through verification and vulnerability discovery.
For example, Hirai first performs a formal verification of a smart contract that is used by the Ethereum Name Service~\cite{hirai2016formal}.\footnote{The Ethereum Name Service is a decentralized service, built on smart contracts, for addressing resources using human-readable names.}
However, this verification proves only one particular property and it involves relatively large amount of manual analysis.
In later work, Hirai defines the complete instruction set of the Ethereum Virtual Machine in Lem, a language that can be compiled for interactive theorem provers~\cite{hirai2017defining}.
Using this definition, certain safety properties can be proven for existing contracts.

Bhargavan et al.\ outline a framework for analyzing and verifying the safety and correctness of Ethereum smart contracts~\cite{bhargavan2016short}.
The framework is built on tools for translating Solidity and Ethereum Virtual Machine bytecode contracts into $F^*$, a functional programming language aimed
at program verification.
Using the~$F^*$ representations, the framework can verify the correctness of the Solidity-to-bytecode compilation as well as detect certain vulnerable patterns. 

Luu et al.\ propose two approaches for mitigating common vulnerabilities in smart contracts~\cite{luu2016making}.
First, they recommend changes to the execution semantics of Ethereum, which eliminate vulnerabilities from the four classes that they identify in their paper.
However, these changes would need to be adopted by all Ethereum clients. 
As a solution that does not require changing Ethereum, they provide a tool called \textsc{Oyente}, which can analyze smart contracts and detect certain security vulnerabilities.

Fr{\"o}wis and B{\"o}hme define a heuristic indicator of control flow immutability to quantify the prevalence of contractual loopholes based on modifying the control flow of Ethereum contracts~\cite{frowis2017code}.
Based on an evaluation of all the contracts deployed on Ethereum, they find that two out of five contracts require trust in at least one third party.

\section{Defining Smart Contracts as FSMs}
\label{sec:FSM}

Let us consider a blind auction (similar to the one presented in \cite{solidityExample}), in which a bidder does not send her actual bid but only a hashed version of it. The bidder is also required make a deposit---which does not need to be equal to her actual bid---to prevent the bidder from not sending the money after she has won the auction. A deposit is considered valid if its value is higher than or equal to the actual bid. We consider that a blind auction has four main \textit{states}: 
\begin{enumerate}
\item \texttt{AcceptingBlindedBids}, in which blind bids and deposits are accepted by the contract;
\item \texttt{RevealingBids}, in which bidders reveal their bids, i.e., they send their actual bids and the contract checks whether the hash value is the same as the one provided during the \texttt{AcceptingBlindedBids} state and whether sufficient deposit has been provided;
\item \texttt{Finished}, in which the highest bid wins the auction. Bidders can withdraw their deposits except for the winner, who can withdraw only the difference between her deposit and bid;
\item \texttt{Canceled}, in which bidders can retract bids and withdraw their deposits.
\end{enumerate}

Our approach relies on the following observations. Smart contracts have \textit{states} (e.g., \texttt{AcceptingBlindedBids}, \texttt{RevealingBids}). Furthermore, contracts provide functions that allow other entities (e.g., contracts or users) to invoke \textit{actions} and change the state of the smart contracts. Thus, smart contracts can be naturally represented by FSMs~\cite{solidityPatterns}. An FSM has a finite set of states and a finite set of transitions between these states. A transition forces a contract to take a set of actions if the associated conditions, which are called the \textit{guards} of the transition, are satisfied. Since such states and transitions have intuitive meaning for developers, representing contracts as FSMs provides an adequate level of abstraction for reasoning about their behavior.

\begin{figure} [t]
\centering
\includegraphics[scale=1]{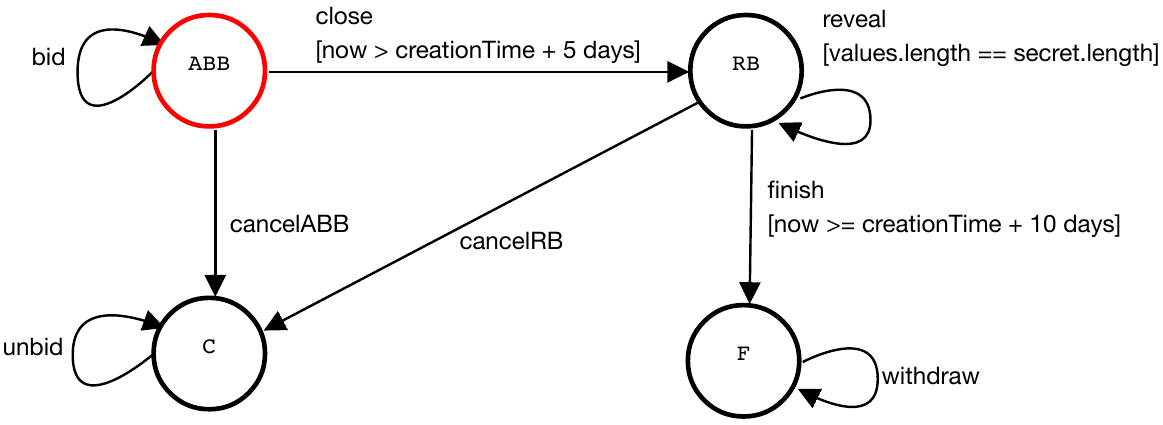}
\caption{Example FSM for blinded auctions.}
\label{fig:blauction}
\end{figure}

Figure \ref{fig:blauction} presents the blind auction example in the form of an FSM. For simplicity, we have abbreviated \texttt{AcceptingBlindedBids}, \texttt{RevealingBids}, \texttt{Finished}, and \texttt{Canceled} to \texttt{ABB}, \texttt{RB}, \texttt{F}, and \texttt{C}, respectively. \texttt{ABB} is the initial state of the FSM. Each transition (e.g., \texttt{bid}, \texttt{reveal}, \texttt{cancel}) is associated to a set of actions that a user can perform during the blind auction. For instance, a bidder can execute the \texttt{bid} transition at the \texttt{ABB} state to send a blind bid and a deposit value. Similarly, a user can execute the \texttt{close} transition, which signals the end of the bidding period, if the associated guard \texttt{now >= creationTime + 5 days} evaluates to true. To differentiate transition names from guards, we use square brackets for the latter. A bidder can reveal her bids by executing the \texttt{reveal} transition. The \texttt{finish} transition signals the completion of the auction, while the \texttt{cancelABB} and \texttt{cancelRB} transitions signal the cancellation of the auction. Finally, the \texttt{unbid} and \texttt{withdraw} transitions can be executed by the bidders to withdraw their deposits. 
For ease of presentation, we omit from Figure \ref{fig:blauction} the actions that correspond to each transition. For instance, during the execution of the \texttt{withdraw} transition, the following action is performed \texttt{amount = pendingReturns[msg.sender]}.  

Guards are based on a set of variables, e.g., \texttt{creationTime}, \texttt{values}, and actions are also based on a set of variables, e.g., \texttt{amount}. These variable sets store data, that can be of type:
\begin{compactitem}
\item \textit{contract data}, which is stored within the contract;
\item \textit{input data}, which is received as transition input;
\item \textit{output data}, which is returned as transition output.
\end{compactitem}
%
We denote by $C$, $I$, and $O$ the three sets of the contract, input, and output variables of a smart contract. We additionally denote:
\begin{align*}
  	&\mathbb{B}[C, I]\,, 
  \text{ the set of Boolean predicates on contract and input variables;}
 \\
&\mathbb{E}[C, I, O]\,,
 \text{ the set of statements that can be defined by the full Solidity syntax.
     }
\end{align*}
%
Notice that $\mathbb{E}[C, I, O]$ represents the set of actions of all transitions. Next, we formally define a contract as an FSM.
\begin{definition}
A Smart Contract is a tuple $(S, s_0, C, I, O, \rightarrow)$, where:
\begin{compactitem}
\item $S$ is a finite set of states;
\item $s_0 \in S$ is the initial state;
\item $C$, $I$, and $O$ are disjoint finite sets of, respectively, contract, input, and output variables;
\item $\rightarrow\, \subseteq S \times \mathcal{G}\times \mathcal{F} \times S$ is a transition relation, where:
  \begin{itemize}
  \item $\mathcal{G} = \mathbb{B}[C, I]$ is a 
  set of guards;
  \item $\mathcal{F}$ is a set of action sets, i.e., a set of all ordered powersets of $\mathbb{E}[C, I, O]$ 
%
  \end{itemize}
\end{compactitem}
\end{definition}


\section{FSM-to-Solidity Transformation}
\label{sec:transformation}

\newcommand{\code}[1]{\textcolor{blue}{\texttt{#1}}}
\newcommand{\lbl}[1]{\textit{#1}}
\newcommand{\ind}{~~~}

To automatically generate a contract using our framework, developers can provide the corresponding FSM in a graphical form. 
Each transition of the FSM is implemented as a Solidity function, where an element of $\mathcal{G}$ and a list of statements from $\mathcal{F}$ form the body. The input $I$ and output $C$ variables correspond to the arguments and the \texttt{return} values, respectively, of these functions.
In this section, we describe the basic transformation formally, while in Section~\ref{sec:plugins}, we present a set of extensions, which we call \textit{plugins}.



First, let us list the input that must be provided by the developer:
\begin{compactitem}
\item $name$: name of the FSM; 
\item $S$: set of states;
\item $s_0 \in S$: initial state;
\item $C$: set of contract variables;
\item for each contract variable $c \in C$, $access(c) \in \{\texttt{public}, \texttt{private}\}$: visibility of the variable;
\item $\rightarrow$: set of transitions;
\item for each transition $t \in \rightarrow$: 
\begin{compactitem}
\item $t^{name}$: name of the transition; 
\item $t^{guards} \in \mathcal{G}$: guard conditions of the transition;
\item $t^{input} \subseteq I$: input variables (i.e., parameters) of the transition; 
\item $t^{statements} \in \mathcal{F}$: statements of the transition; 
\item $t^{output} \subseteq O$: output (i.e., return values) of the transition; 
\item $t^{from} \in S$: previous state;
\item $t^{to} \in S$: next state;
\item 
$t^{tags} \subseteq \{\lbl{payable}, \lbl{admin}, \lbl{event}\}$: set of transition properties specified by the developer (note that without plugins, only \lbl{payable} is supported);
\end{compactitem}
\item $\mathcal{T}^{custom}$: set of complex types defined by the developer in the form of \texttt{struct}s.
\end{compactitem}
For any variable $v \in C \cup I \cup O$, we let $type(v) \in \mathcal{T}$ denote the domain of the variable,
where $\mathcal{T}$ denotes the set of all built-in Solidity types and developer-defined \texttt{struct} types. 

We use $\code{fixed-width}$ font 
for the output generated by the transformation, and $\lbl{italic}$ font for elements that are replaced with input or specified later.
An FSM is transformed into a Solidity contract as follows:
\begin{align*}
\lbl{Contract} ::= ~&\code{contract } name \code{ \{} \\
& \ind \lbl{StatesDefinition} \\
& \ind \code{uint private creationTime = now;} \\
& \ind \lbl{VariablesDefinition} \\ 
& \ind \lbl{Plugins} \\
& \ind \lbl{Transition}(t_1) \\
& \ind \ldots \\
& \ind \lbl{Transition}(t_{|\rightarrow|}) \\
& \code{\}}
\end{align*}
where $\{ t_1, \ldots, t_{|\rightarrow|}\} =~\rightarrow$.
Without any security extensions or design patterns added (see Section~\ref{sec:plugins}), $\lbl{Plugins}$ is empty. Appendix \ref{sec:extensionsExampleAppendix} contains the complete generated code for the blind-auction example presented in Figure \ref{fig:blauction} (with the locking and transition-counter security-extension plugins added).

\begin{align*}
\lbl{StatesDefinition} ::= ~&\code{enum States \{} s_0 \code{,} \ldots \code{,} s_{|S|-1} \code{\}} \\
& \code{States private state = States.} s_0 \code{;}
\end{align*}
where $\{ s_0, \ldots, s_{|S|-1} \} = S$.

\begin{example}
The following snippet of Solidity code presents the $\lbl{StatesDefinition}$ code generated for the blind auction example (see Figure \ref{fig:blauction}).
\begin{lstlisting}[style=customjavaexample]
    enum States {
        ABB,
        RB,
        F,
        C
    }
    States private state = States.ABB;
\end{lstlisting}
\end{example}

\begin{align*}
\lbl{VariablesDefinition} ::= ~& \mathcal{T}^{custom} \\
& type(c_1) ~ access(c_1) ~ c_1 \code{;} \\
& \ldots \\
& type(c_{|C|}) ~ access(c_{|C|}) ~ c_{|C|} \code{;} 
\end{align*}
where $\{ c_1, \ldots, c_{|C|} \} = C$. 

\begin{example}
The following snippet of Solidity code presents the $\lbl{VariablesDefinition}$ code of the blind auction example (see Figure \ref{fig:blauction}).
\begin{lstlisting}[style=customjavaexample]
    struct Bid {
        bytes32 blindedBid;
        uint deposit;
    }
    mapping(address => Bid[]) private bids;
    mapping(address => uint) private pendingReturns;
    address private highestBidder;
    uint private highestBid;
\end{lstlisting}
\end{example}
\begin{align*}
\lbl{Transition}(t) ::= ~&\code{function } t^{name} 
\code{(} type(i_1)~ i_1\code{, } \ldots\code{, } type(i_{|t^{input}|})~ i_{|t^{input}|} \code{)} \\ 
& \ind\ind  \lbl{TransitionPlugins}(t) \\
& \ind\ind \lbl{Payable}(t) ~ \lbl{Returns}(t) \code{ \{} \\
& \ind \code{require(state == States.} t^{from} \code{);} \\
& \ind \lbl{Guards}(t) \\
& \ind \lbl{Statements}(t) \\
& \ind \code{state = States.}t^{to}\code{;} \\
& \code{\}} 
\end{align*}
where $\left\{i_1, \ldots, i_{|t^{input}|}\right\} = t^{input}$. 
Without any security extensions or design patterns (see Section~\ref{sec:plugins}), $\lbl{TransitionPlugins}(t)$ is empty, similar to $\lbl{Plugins}$.
If $\lbl{payable} \in t^{tags}$, then $\lbl{Payable}(t) = \code{payable}$; otherwise, it is empty. 
If $t^{to} = t^{from}$ then the line $\code{state = States.}t^{to}\code{;}$ is not generated.

If $t^{output} = \emptyset$, then $\lbl{Returns}(t)$ is empty. Otherwise, it is as follows:
\begin{align*}
\lbl{Returns}(t) ::= \code{returns (}  type(o_1)~ o_1 \code{, } \ldots \code{, } type(o_{|t^{output}|})~ o_{|t^{output}|} \code{)}
\end{align*}
where $\left\{ o_1, \ldots, o_{|t^{output}|} \right\} = t^{output}$. 

Further,
\begin{align*}
\lbl{Guards}(t) ::= ~&\code{require( (} g_1 \code{) \&\& (} g_2 \code{) \&\& } \ldots \code{ \&\& (} g_{|t^{guards}|} \code{) );} \\
\lbl{Statements}(t) ::= ~& a_1 \\ 
& \ldots \\
& a_{|t^{statements}|}  
\end{align*}
where $\{ g_1, \ldots, g_{|t^{guards}|} \} = t^{guards}$
and $\{ a_1, \ldots, a_{|t^{statements}|} \} = t^{statements}$. 

\begin{example}
The following snippet of Solidity code shows the generated \texttt{bid} transition (see Figure \ref{fig:blauction}). The \texttt{bid} transition does not have any guards and the state of the FSM does not change, i.e., it remains \texttt{ABB} after the execution of the transition.
\begin{lstlisting}[style=customjavaexample]
    // Transition bid
    function bid(bytes32 blindedBid)
      payable
    {
        require(state == States.ABB);
        //Actions
        bids[msg.sender].push(Bid({
            blindedBid: blindedBid,
            deposit: msg.value
        }));
    }
\end{lstlisting}
\end{example}

\begin{example}
The following snippet of Solidity code shows the generated \texttt{close} transition (see Figure \ref{fig:blauction}). The \texttt{close} transition does not have any associated actions but the state of the FSM changes from \texttt{ABB} to \texttt{RB} after the execution of the transition.
\begin{lstlisting}[style=customjavaexample]
    // Transition close
    function close()
    {
        require(state == States.ABB);
        //Guards
        require(now >= creationTime + 5 days);
        //State change
        state = States.RB;
    }
\end{lstlisting}
\end{example}

\section{Security Extensions and Patterns}
\label{sec:plugins}

Building on the FSM model and the FSM-to-Solidity transformation introduced in the previous sections, we next provide extensions and patterns for enhancing the security and functionality of contracts.
These extensions and patterns are implemented as plugins, which are appended to the $\lbl{Plugins}$ and $\lbl{TransitionPlugins}$ elements.
Developers can easily add plugins to a contract (or some of its transitions) using our tool, without writing code manually.\footnote{Please note that we introduce an additional plugin in Appendix~\ref{sec:events}, which we omitted from the main text due to lack of space.}

\subsection{Locking}
\label{sec:locking}

To prevent reentrancy vulnerabilities, we provide a security plugin for locking the smart contract.
\footnote{\url{http://solidity.readthedocs.io/en/develop/contracts.html?highlight=mutex\#function-modifiers}}
The locking feature eliminates reentrancy vulnerabilities in a ``foolproof'' manner:  
functions within the contract cannot be nested within each other in any way.

\subsubsection{Implementation}
If the locking plugin is enabled, then
\begin{align*}
\lbl{Plugins} ~{+\!\!=} ~&\code{bool private locked = false;} \\
&\code{modifier locking \{} \\
& \ind \code{require(!locked);} \\
& \ind \code{locked = true;} \\
& \ind \code{\char`_;} \\
& \ind \code{locked = false;} \\
&\code{\}}
\end{align*}
and for every transition $t$,
\begin{align*}
\lbl{TransitionPlugins}(t) ~{+\!\!=} ~&\code{locking}
\end{align*}
Before a transition is executed, the $\texttt{locking}$ modifier first checks if the contract is locked.
If it is not locked, then the modifier locks it, executes the transition, and unlocks it after the transition has finished.
Note that the $\texttt{locking}$ plugin must be applied before the other plugins so that it can prevent reentrancy vulnerabilities in the other plugins. Our tool always applies plugins in the correct order. 

\subsection{Transition Counter}
\label{sec:transCounter}

Recall from Section~\ref{sec:commonVuln} that the state and the values of the variables stored in an Ethereum contract may be unpredictable: when a user invokes a function (i.e., transition in an FSM), she cannot be sure that the contract does not change in some way before the function is actually executed.
This issue has been referred to as ``transaction-ordering dependence''~\cite{luu2016making} and ``unpredictable state''~\cite{atzei2017survey}, and it can lead to various security issues.
Furthermore, it is rather difficult to prevent since multiple users may invoke functions at the same time, and these function invocations might be executed in any order.

We provide a plugin that can prevent unpredictable-state vulnerabilities by enforcing a strict ordering on function executions.
The plugin expects a transition number in every function as a parameter (i.e., as a transition input variable) and ensures that the number is incremented by one for each function execution.
As a result, when a user invokes a function with the next transition number in sequence, she can be sure that the function is executed before any other state changes can take place (or that the function is not executed).

\subsubsection{Implementation}
If the transition counter plugin is enabled, then
\begin{align*}
\lbl{Plugins} ~{+\!\!=} ~&\code{uint private transitionCounter = 0;} \\
&\code{modifier transitionCounting(uint nextTransitionNumber) \{} \\
& \ind \code{require(nextTransitionNumber == transitionCounter);} \\
& \ind \code{transitionCounter += 1;} \\
& \ind \code{\char`_;} \\
&\code{\}}
\end{align*}
and for every transition $t$,
\begin{align*}
\lbl{TransitionPlugins}(t) ~{+\!\!=} ~&\code{transitionCounting(nextTransitionNumber)}
\end{align*}
\todo{Clarify!}
Note that due to the inclusion of the above modifier, $t^{input}$---and hence the parameter list of every function implementing a transition--- includes the parameter $\texttt{nextTransitionNumber}$ of type $\texttt{uint}$.

\subsection{Automatic Timed Transitions}
\label{sec:timedTrans}

Next, we provide a plugin for implementing time-constraint patterns.
We first need to extend our FSM model: a Smart Contract with Timed Transitions is a tuple $C = (S, s_0, C, I, O, \rightarrow, \overset{T}{\rightarrow})$, where $\overset{T}{\rightarrow} \subseteq S \times \mathcal{G}_T \times \mathbb{N} \times \mathcal{F}_T \times S$ is a timed transition relation such that
\begin{compactitem}
\item $\mathcal{G}_T = \mathbb{B}[C]$ is a set of guards (without any input data),
\item $\mathbb{N}$ is the set of natural numbers, which is used to specify the time of the transition in seconds,
\item $\mathcal{F}_T$ is a set of action sets, i.e., a set of all ordered powerset of ${\mathbb{E}[C]}$.
\end{compactitem}
Notice that timed transitions are similar to non-timed transitions, but 1)~their guards and assignments do not use input or output data and 2)~they include a number specifying the transition time.

We implement timed transitions as a modifier that is applied to every function.
When a transition is invoked, the modifier checks whether any timed transitions must be executed before the invoked transition is executed. If so, the modifier executes the timed transitions before the invoked transition.

Writing such modifiers for automatic timed transitions manually may lead to vulnerabilities.
For example, a developer might forget to add a modifier to a function, which enables malicious users to invoke functions without the contract progressing to the correct state (e.g., place bids in an auction even though the auction should have already been closed due to a time limit).

\subsubsection{Implementation}
For every timed transition $tt \in \overset{T}{\rightarrow}$, the developer specifies a time $tt^{time} \in \mathbb{N}$ at which the transition will automatically happen (given that the guard condition is met).
This time is measured in the number of seconds elapsed since the creation (i.e., instantiation) of the contract.
We let $tt_1, tt_2, \ldots$ denote the list of timed transitions in ascending order based on their specified times. When the plugin is enabled,
\begin{align*}
\lbl{Plugins} ~{+\!\!=} ~&\code{modifier timedTransitions \{} \\
& \ind \lbl{TimedTransition}(tt_1) \\
& \ind \lbl{TimedTransition}(tt_2) \\
& \ind \ldots \\
& \ind \code{\char`_;} \\
& \code{\}} 
\end{align*}
where
\begin{align*}
\lbl{TimedTransition}(t) ~{::=} ~&\code{if ((state == States.}t^{from}\code{)} \\
& \ind\ind \code{\&\& (now >= creationTime + } t^{time}\code{) } \\
& \ind\ind \code{\&\& (} \lbl{Guard}(t)\code{)) \{} \\
& \ind \lbl{Statements}(t) \\
& \ind \code{state = States.}t^{to}\code{;} \\
& \code{\}}
\end{align*}

Finally, for every non-timed transition $t \in \rightarrow$, let
\begin{align*}
\lbl{TransitionPlugins}(t) ~{+\!\!=} ~&\code{timedTransitions}
\end{align*}

\subsection{Access Control}
\label{sec:accessContr}

In many contracts, access to certain transitions (i.e., functions) needs to be controlled and restricted.
\footnote{\url{http://solidity.readthedocs.io/en/develop/common-patterns.html\#restricting-access}}
For example, any user can participate in a typical blind auction by submitting a bid, but only the creator should be able to cancel the auction. 
To facilitate the enforcement of such constraints, we provide a plugin that 1) manages a list of administrators at runtime (identified by their addresses) and 2) enables developers to forbid non-administrators from accessing certain functions.
This plugin implements management functions (\texttt{addAdmin}, \texttt{removeAdmin}) for only one privileged group, but it could easily be extended to support more fine-grained access control.

Due to space limitation, the implementation of the access control plugin is in Appendix~\ref{sec:accessContrImpl}.

\section{The FSolidM Tool}
\label{sec:tool}

We present the FSolidM tool, which is build on top of WebGME~\cite{maroti2014next}, a web-based, collaborative, versioned, model editing framework. FSolidM enables collaboration between multiple users during the development of smart contracts. Changes in FSolidM are committed and versioned, which enables branching, merging, and viewing the history of a contract. 
\ifAnonymous
FSolidM is open-source~\footnote{\url{https://github.com/FSolidM/smart-contracts}}.
\else
FSolidM is open-source~\footnote{\url{https://github.com/anmavrid/smart-contracts}} and available online~\footnote{\url{https://cps-vo.org/group/SmartContracts}}.
\fi

\begin{figure}[h]
\centering
\includegraphics[width=\textwidth]{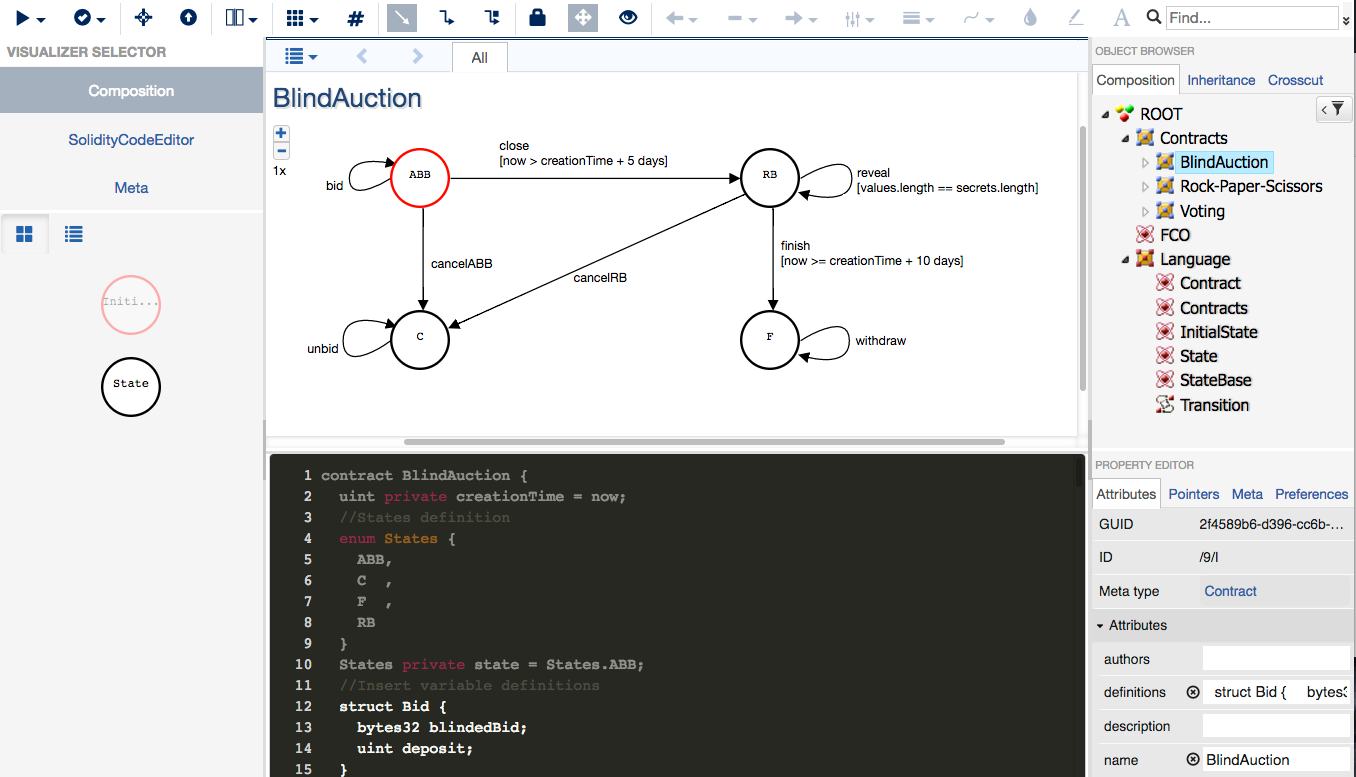}
\caption{The graphical and code editors provided by FSolidM.}
\label{fig:editors}
\end{figure}

To use FSolidM, a developer must provide some input (see Section \ref{sec:transformation}). To do so, the developer can use the graphical editor of FSolidM to specify the states, transitions, guards, etc. of a contract. The full input of the smart-contract code generator can be  defined entirely through the FSolidM graphical editor. For the convenience of the developers, we have also implemented a Solidity code editor, since part of the input e.g., variable definitions and function statements, might be easier to directly write in a code editor. Figure \ref{fig:editors} shows the two editors of the tool. We have integrated a Solidity parser\footnote{\url{https://github.com/ConsenSys/solidity-parser}} to check the syntax of the Solidity code that is given as input by the developers. 

The FSolidM code editor cannot be used to completely specify the required input. Notice that in Figure \ref{fig:editors}, parts of the code shown in the code editor are darker (lines 1-10) than other parts (lines 12-15). The darker lines of code include code that was generated from the FSM model defined in the graphical editor and are locked---cannot be altered in the code editor. The non-dark parts indicate code that was directly specified in the code editor. 

FSolidM provides mechanisms for checking if the FSM is correctly specified (e.g., whether an initial state exists or not). FSolidM notifies developers of errors and provides links to the erroneous nodes of the model (e.g., a transition or a guard). Additionally, FSolidM provides an FSM-to-Solidity code generator and mechanisms for easily integrating the plugins introduced in Section \ref{sec:plugins}. 
\ifToolDetails
We present the FSolidM tool in greater detail in Appendix~\ref{sec:appendixtool}.
\fi


\subsection{Numerical Results on Computational Cost}
\label{sec:numerical}

Plugins not only enhance security but also increase the computational cost of transitions.
Since users must pay a relatively high price for computation performed on the public Ethereum platform, the computational cost of plugins is a critical question.
Here, we measure and compare the computational cost of transitions in our blind-auction contract without and with the locking and transition counter plugins.
We focus on these security feature plugins because they introduce overhead, while the design pattern plugins introduce useful functionality.

For this experiment, we use Solidity compiler version 0.4.17 with optimizations enabled.
In all cases, we quantify computational cost of a transition as the gas cost of an Ethereum transaction that invokes the function implementing the transition.\footnote{Gas measures the cost of executing computation on the Ethereum platform.}
The cost of deploying our smart contract was 504,672 gas without any plugins, 577,514 gas with locking plugin, 562,800 gas with transition counter plugin, and 637,518 gas with both plugins.\footnote{At the time of writing, this cost of deployment was well below \$1 (if the deployment does not need to be prioritized).}

\begin{figure}[t]
\centering
\begin{tikzpicture}
\begin{axis}[
    width=1.05\textwidth,
    height=0.5\textwidth,
    bar width=0.018\textwidth,
    ybar,
    ymin=0,
    ylabel={Transaction cost [gas]},
    legend pos=north east,
    symbolic x coords={{bid}, {cancelABB}, {unbid}, {close}, {reveal}, {finish}, {withdraw}},
    xtick=data,
    xticklabel style={font=\small\tt},
    ]
    \addplot coordinates { 
        (bid, 58249)        
        (cancelABB, 42059)  
        (unbid, 19735)
        (close, 42162)
        (reveal, 65729)
        (finish, 27239)
        (withdraw, 20290)
    };
    \addlegendentry{without plugins};
    \addplot coordinates { 
        (bid, 68917)        
        (cancelABB, 52727)
        (unbid, 30406)
        (close, 52830)
        (reveal, 76415)
        (finish, 37913)
        (withdraw, 30961)
    };
    \addlegendentry{with locking};
    \addplot coordinates { 
        (bid, 63924)        
        (cancelABB, 47661)
        (unbid, 25406)
        (close, 47764)
        (reveal, 71390)
        (finish, 32891)
        (withdraw, 25961)
    };
    \addlegendentry{with counter};
    \addplot coordinates { 
        (bid, 74607)        
        (cancelABB, 58329)
        (unbid, 36074)
        (close, 58432)
        (reveal, 82067)
        (finish, 43559)
        (withdraw, 36629)
    };
    \addlegendentry{with both};
\end{axis}
\end{tikzpicture}
\caption{Transaction costs in gas without plugins (\textcolor{blue}{\bf blue}), with locking plugin (\textcolor{red}{\bf red}), with transition counter plugin (\textcolor{brown}{\bf brown}), and with both plugins (\textcolor{darkgray}{\bf dark gray}).}
\label{fig:transCosts}
\end{figure}
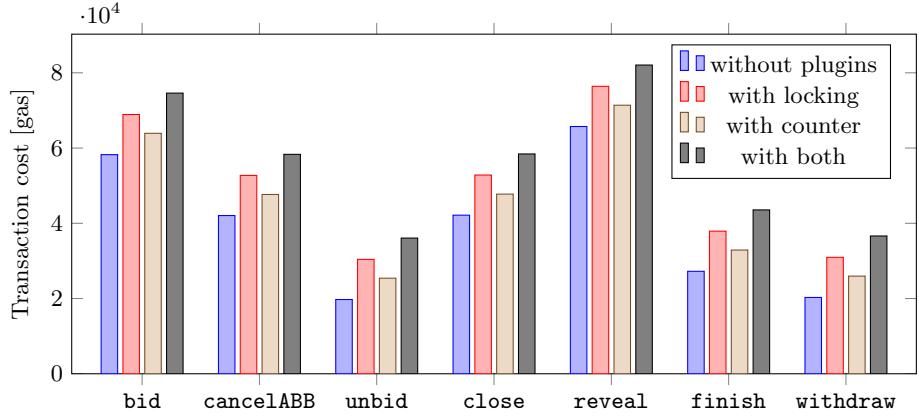

Figure~\ref{fig:transCosts} (and Table~\ref{tab:transCosts} in Appendix~\ref{sec:numericalTable}) shows the gas cost of each transition for all four combinations of the two plugins. 
We make two key observations.
First, \emph{computational overhead is almost constant} for both plugins and also for their combination.
For example, the computational overhead introduced by locking varies between 10,668 and 10,686 gas.
For the simplest transition, \texttt{unbid}, this constitutes a 54\%  increase in computational cost,
while for the most complex transition, \texttt{reveal}, the increase is 16\%.
Second, the \emph{computational overhead of the two plugins is additive}. The increase in computational cost for enabling locking, transition counter, and both are around 10,672 gas, 5,648 gas, and 16,319 gas, respectively.

\section{Conclusion and Future Work}
\label{sec:concl}

Distributed computing platforms with smart-contract functionality are envisioned to have a significant technological and economic impact in the future.
However, if we are to avoid an equally significant risk of security incidents,
we must ensure that smart contracts are secure.
While previous research efforts focused on identifying vulnerabilities in existing contracts, we explored a different avenue by proposing and implementing a novel framework for creating secure smart contracts.
We introduced a formal, FSM based model of smart contracts.
Based on this model, we implemented a graphical editor for designing contracts as FSMs and an automatic code generator.
We also provided a set of plugins that developers can add to their contracts.
Two of these plugins, \emph{locking} and \emph{transition counter}, implement security features for preventing common vulnerabilities (i.e., reentrancy and unpredictable state).
The other two plugins, \emph{automatic timed transitions} and \emph{access control}, implement common design patterns to facilitate the development of correct contracts with complex functionality.


We plan to extend our framework in multiple directions.
First, we will introduce a number of plugins, implementing various security features and design patterns. 
We will provide security plugins for all the vulnerability types identified in~\cite{atzei2017survey} that can be addressed on the level of Solidity code.
\todo{Reference the events plugins as an example?}
We will also provide plugins implementing the most popular design patterns surveyed in~\cite{bartoletti2017empirical}. 

Second, we will integrate verification tools~\cite{bensalem2009d,cavada2014nuxmv} and correctness-by-design techniques~\cite{mavridou2016satellite} into our framework.
This will enable developers to easily verify the security and safety properties of their contracts.
For example, developers will be able to verify if a malicious user could lead a contract into a deadlocked state.
Recall that deadlocks present a serious issue since it may be impossible to recover the functionality or assets of a deadlocked contract.

Third, we will enable developers to model and verify multiple interacting contracts as a set of interacting FSMs.
By verifying multiple contracts together, developers will be able to identify a wider range of issues.
For example, a set of interacting contracts may get stuck in a deadlock even if the individual contracts are deadlock free.

\bibliographystyle{splncs}
\bibliography{references}

\allowdisplaybreaks

\appendix

\section{Implementation of the Access Control Plugin}
\label{sec:accessContrImpl}


Here, we describe the implementation of the access control plugin, which is introduced in Section~\ref{sec:accessContr}.

If the access control plugin is enabled, then
\begin{align*}
\lbl{Plugins} ~{+\!\!=} ~&\code{mapping(address => bool) private isAdmin;} \\
&\code{uint private numAdmins = 1;} \\
& \\
& \code{function } name \code{() \{} \\
& \ind \code{isAdmin[msg.sender] = true;} \\
& \code{\}} \\
& \\
& \code{modifier onlyAdmin \{} \\
& \ind \code{require(isAdmin[msg.sender]);} \\
& \ind \code{\char`_;} \\
& \code{\}} \\
& \\
& \code{function addAdmin(address admin) onlyAdmin \{} \\
& \ind \code{require(!isAdmin[admin]);} \\
& \ind \code{isAdmin[admin] = true;} \\
& \ind \code{numAdmins += 1;} \\
& \code{\}} \\
& \\
& \code{function removeAdmin(address admin) onlyAdmin \{} \\
& \ind \code{require(isAdmin[admin]);} \\
& \ind \code{require(numAdmins > 1);} \\
& \ind \code{isAdmin[admin] = false;} \\
& \ind \code{numAdmins -= 1;} \\
& \code{\}}
\end{align*}
For transitions $t$ such that $\lbl{admin} \in t^{tags}$ (i.e., transitions that are tagged ``only admin'' by the developer), 
\begin{align*}
\lbl{TransitionPlugins}(t) ~{+\!\!=} ~&\code{onlyAdmin}
\end{align*}


\section{Event Plugin}
\label{sec:events}

In this section, we introduce an additional plugin, which developers can use to notify users of transition executions.
The \emph{event plugin} uses the \texttt{event} feature of Solidity, which provides a convenient interface to the Ethereum logging facilities.
If this plugin is enabled, transitions tagged with $\lbl{event}$ emit a Solidity event after they are executed.
Ethereum clients can listen to these events, allowing them to be notified when a tagged transition is executed on the platform.

\subsubsection{Implementation}

If the event plugin is enabled, then
\begin{align*}
\lbl{Plugins} ~{+\!\!=} ~&\lbl{TransitionEvent}(t_1) \\
~&\lbl{TransitionEvent}(t_2) \\
~&\ldots 
\end{align*}
where $\{t_1, t_2, \ldots\}$ is the set of transitions with the tag $\lbl{event}$.
\begin{align*}
\lbl{TransitionEvent}(t) ::= ~&\code{event Event}t^{name}\code{;} \\
&\code{modifier event}t^{name} \code{ \{} \\
& \ind \code{\char`_;} \\
& \ind \code{Event}t^{name}\code{();} \\
&\code{\}}
\end{align*}

For every transition $t$ such that $\lbl{event} \in t^{tags}$ (i.e., transitions that are tagged to emit an event),
\begin{align*}
\lbl{TransitionPlugins}(t) ~{+\!\!=} ~&\code{event}t^{name}
\end{align*}

\section{Blind Auction: Security Extensions}
\label{sec:extensionsExampleAppendix}

We list below the complete code of the blind auction example, as generated by the FSolidM tool. In the following code, the \emph{locking} and the \emph{transition counter} plugins have been applied.

\begin{lstlisting}[style=customjava]
contract BlindAuction{
    //States definition
    enum States {
        ABB,
        RB,
        F,
        C
    }
    States private state = States.ABB;
    
    //Variables definition
    struct Bid {
        bytes32 blindedBid;
        uint deposit;
    }
    mapping(address => Bid[]) private bids;
    mapping(address => uint) private pendingReturns;
    address private highestBidder;
    uint private highestBid;
    uint private creationTime = now;

    //Locking
    bool private locked = false;
    modifier locking {
        require(!locked);
        locked = true;
        _;
        locked = false;
    }
    
    //Transition counter
    uint private transitionCounter = 0;
    modifier transitionCounting(uint nextTransitionNumber) {
        require(nextTransitionNumber == transitionCounter);
        transitionCounter += 1;
        _;
    }
    
    //Transitions
    //Transition bid
    function bid (uint nextTransitionNumber, bytes32 blindedBid)
        payable
        locking
        transitionCounting(nextTransitionNumber)
    {
        require(state == States.ABB);
        //Actions
        bids[msg.sender].push(Bid({
            blindedBid: blindedBid,
            deposit: msg.value
        }));
        pendingReturns[msg.sender] += msg.value;
    }
    
    //Transition close
    function close(uint nextTransitionNumber)
        locking
        transitionCounting(nextTransitionNumber)
    {
        require(state == States.ABB);
        //Guards
        require(now >= creationTime + 5 days);
        //State change
        state = States.RB;
    }
    
    //Transition reveal
    function reveal(uint nextTransitionNumber, uint[] values, bytes32[] secrets)
        locking
        transitionCounting(nextTransitionNumber)
    {
        require(state == States.RB);
        //Guards
        require(values.length == secrets.length);
        //Actions
        for (uint i = 0; i < (bids[msg.sender].length < values.length ? 
        bids[msg.sender].length : values.length); i++) {
            var bid = bids[msg.sender][i];
            var (value, secret) = (values[i], secrets[i]);
            if (bid.blindedBid != keccak256(value, secret)) {
                // Do not add to refund value.
                continue;
            }
            if (bid.deposit >= value && value > highestBid) {
                    highestBid = value;
                    highestBidder = msg.sender;
            }
        }
    }
    
    //Transition finish
    function finish(uint nextTransitionNumber)
        locking
        transitionCounting(nextTransitionNumber)
    {
        require(state == States.RB);
        //Guards
        require(now >= creationTime + 10 days);
        //State change
        state = States.F;
    }
    
    //Transition cancelABB
    function cancelABB(uint nextTransitionNumber)
        locking
        transitionCounting(nextTransitionNumber)
    {
        require(state == States.ABB);
        //State change
        state = States.C;
    }
    
    //Transition cancelRB
    function cancelRB(uint nextTransitionNumber)
        locking
        transitionCounting(nextTransitionNumber)
    {
        require(state == States.RB);
        //State change
        state = States.C;
    }
    
    //Transition withdraw
    function withdraw(uint nextTransitionNumber)
        locking
        transitionCounting(nextTransitionNumber)
    { 
        uint amount;
        require(state == States.F);
        //Actions
        amount = pendingReturns[msg.sender];
        if (amount > 0 && msg.sender!= highestBidder) {
            msg.sender.transfer(amount);
            pendingReturns[msg.sender] = 0;
        } else {
            msg.sender.transfer(amount - highestBid);
            pendingReturns[msg.sender] = 0;
        }
    }
    
    //Transition unbid
    function unbid(uint nextTransitionNumber)
        locking
        transitionCounting(nextTransitionNumber)
    {
        uint amount;
        require(state == States.C);
        //Actions
        amount = pendingReturns[msg.sender];
        if (amount > 0) {
            msg.sender.transfer(amount);
            pendingReturns[msg.sender] = 0;
        }
    }   
}

\end{lstlisting}

\section{Numerical Results on Computational Cost -- Table}
\label{sec:numericalTable}

Table~\ref{tab:transCosts} shows the gas cost of each transition for all four combinations of the \emph{locking} and \emph{transition counter} plugins.
Please see Section~\ref{sec:numerical} for a discussion of these results.

\begin{table}
\centering
\caption{Computational Cost of Transitions}
\label{tab:transCosts}
\setlength{\tabcolsep}{0.6em}
\renewcommand{\arraystretch}{1.1}
\begin{tabular}{|l|r|r|r|r|}
\hline
Transition & \multicolumn{4}{c|}{Computational cost [gas]} \\
\cline{2-5}
 & without plugins & with locking & with transition counter & with both \\
\hline
\texttt{bid} & 58,249 & 68,917 & 63,924 & 74,607 \\
\texttt{cancelABB} & 42,059 & 52,727 & 47,661 & 58,329 \\
\texttt{unbid} & 19,735 & 30,406 & 25,406 & 36,074 \\
\texttt{close} & 42,162 & 52,830 & 47,764 & 58,432 \\
\texttt{reveal} & 65,729 & 76,415 & 71,390 & 82,067 \\
\texttt{finish} & 27,239 & 37,913 & 32,891 & 43,559 \\
\texttt{withdraw} & 20,290 & 30,961 & 25,961 & 36,629 \\
\hline
\end{tabular}
\end{table}

\ifToolDetails
\section{Detailed Description of the FSolidM Tool}
\label{sec:appendixtool}

\ifAnonymous
FSolidM is open-source~\footnote{\url{https://github.com/FSolidM/smart-contracts}}.
\else
FSolidM is open-source~\footnote{\url{https://github.com/anmavrid/smart-contracts}} and available online~\footnote{\url{https://cps-vo.org/group/SmartContracts}}.
\fi
Next, we present it in detail.

\subsection{FSolidM Metamodel}

\begin{figure}
\centering
\includegraphics[width=\textwidth]{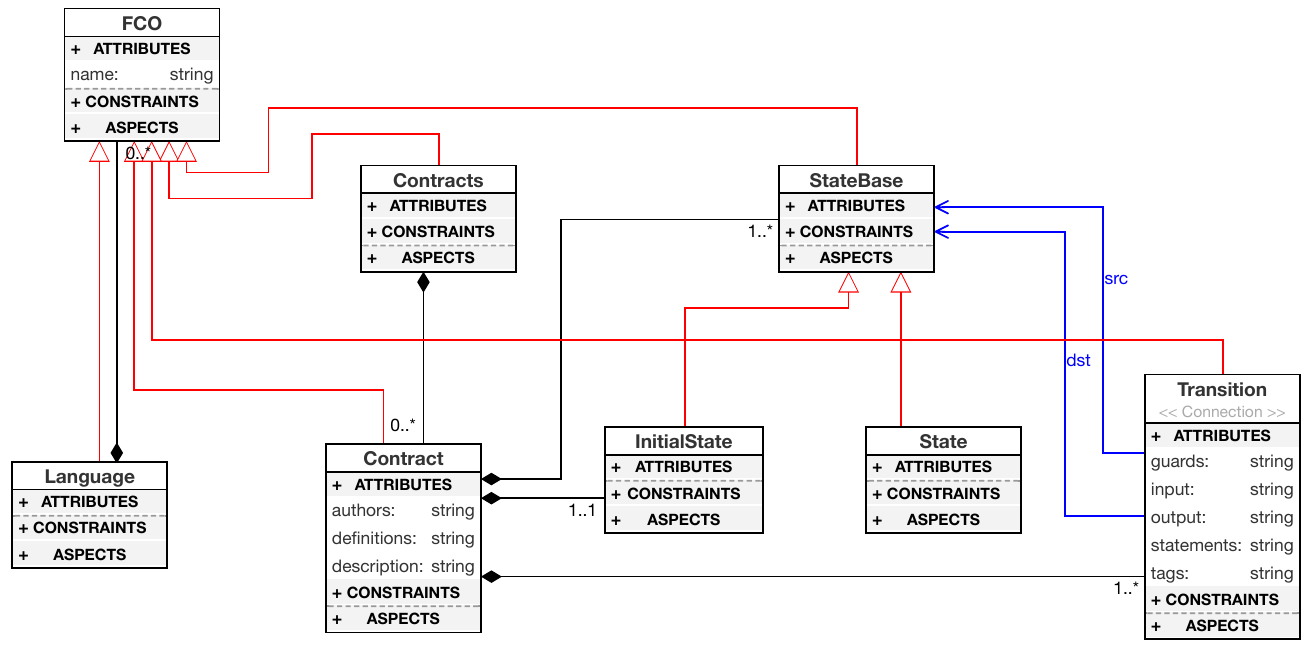}
\caption{The FSolidM metamodel}
\label{fig:meta}
\end{figure}

Figure \ref{fig:meta} shows the FSolidM metamodel as a UML class diagram. The behavior of a \texttt{Contract} is described, as explained in Section \ref{sec:FSM}, by an FSM. The \texttt{State\_Base} concept of the metamodel is abstract and it can be either instantiated by an \texttt{InitialState} or a \texttt{State}. Each contract must have exactly one \texttt{InitialState}, which is enforced by the cardinality of the containment relation. The \texttt{Transition} concept of the metamodel is characterized by six attributes: 1)~\texttt{name} which is inherited from \texttt{FCO} (the base element of our metamodel), 2)~the associated \texttt{guards}, the 3)~\texttt{input} and 4)~\texttt{output} data of the transition, the 5)~\texttt{statements} and, finally, the 6)~\texttt{tags}.

\subsection{Versioning in FSolidM}
Changes in FSolidM are committed and versioned, which enables branching, merging, and viewing the history of a contract. Figure \ref{fig:version} shows the history of the \texttt{master} branch of the blind auction contract.

\begin{figure}
\centering
\includegraphics[width=\textwidth]{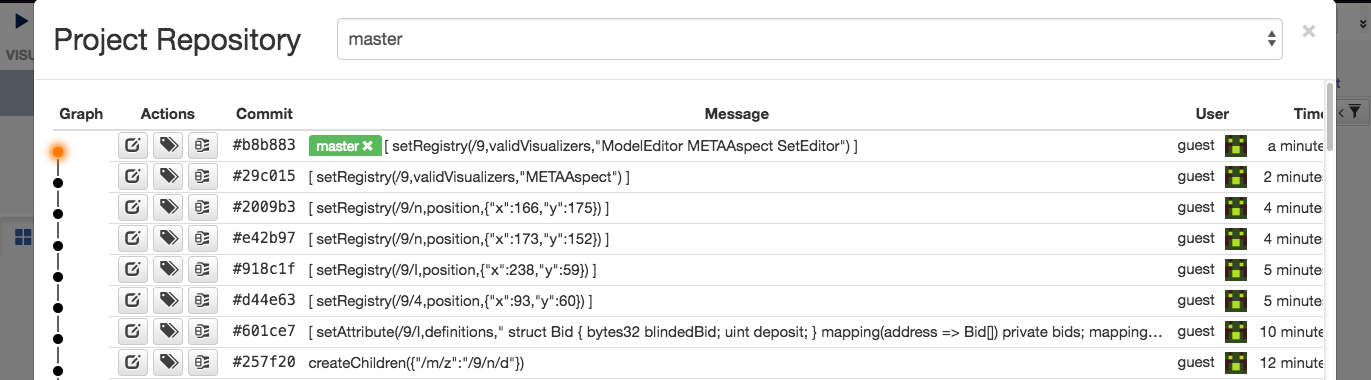}
\caption{Versioning in FSolidM}
\label{fig:version}
\end{figure}

\subsection{The \texttt{SolidityCodeGenerator} and \texttt{AddSecurityPatterns} mechanisms}

To generate Solidity code or apply security extensions and patterns, a developer must (after specifying the required input) click one of the two offered services in the upper left corner of the tool (see Figure \ref{fig:mechanisms}). For instance, if the developer clicks on the \texttt{SolidityCodeGenerator}, then the widget shown in Figure \ref{fig:generator} pops up, and the developer must click on the \texttt{Save \& Run} button to continue with the code generation. Similarly, if the developer clicks on the \texttt{AddSecurityPatterns}, then the widget shown in Figure \ref{fig:patterns} pops up, and the developer can pick the patterns that she wants to apply on her model. 

If the code generation is successful, i.e.,  there are no specification errors in the given input, then, the widget shown in Figure \ref{fig:correct} pops-up. The developer can then click on the generated artifacts to download the generated Solidity contract. If, on the other hand, the code generation is not successful due to incorrect input, then, the widget shown in Figure \ref{fig:wrong} pops-up. As shown in Figure \ref{fig:wrong}, FSolidM lists the errors found in the specification of the contract with detailed explanatory messages. Additionally, FSolidM provides links (through \texttt{Show node}) that redirect the developer to the erroneous nodes of the contract.

\begin{figure} [t]
\centering
\includegraphics[width=\textwidth]{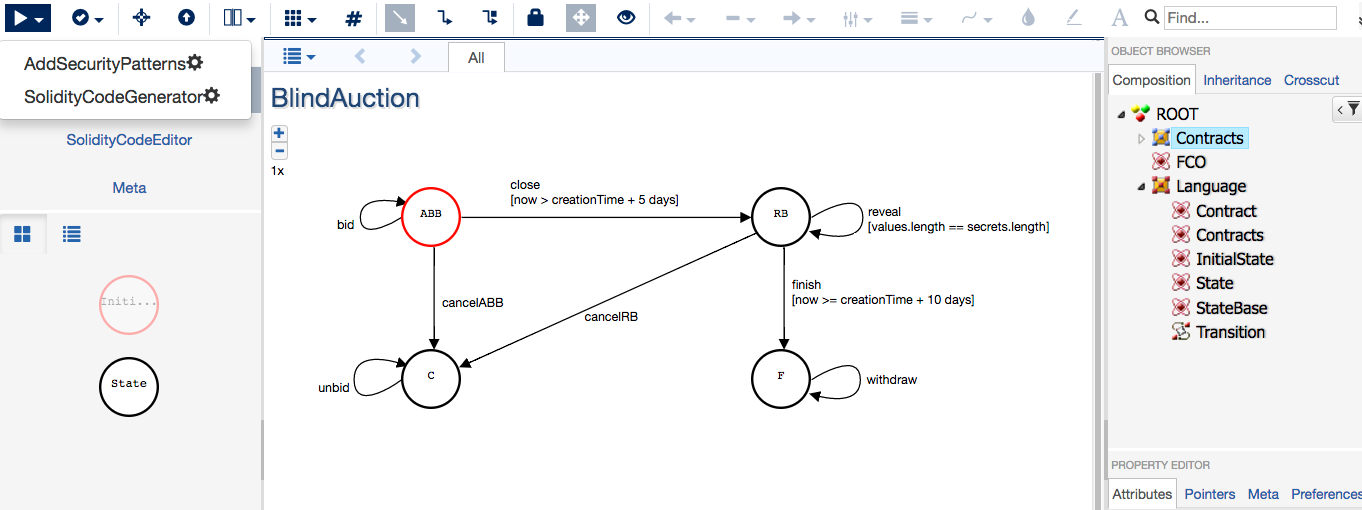}
\caption{Invoking the \texttt{SolidityCodeGenerator} and \texttt{AddSecurityPatterns} mechanisms}
\label{fig:mechanisms}
\end{figure}

\begin{figure} [t]
\centering
\includegraphics[width=\textwidth]{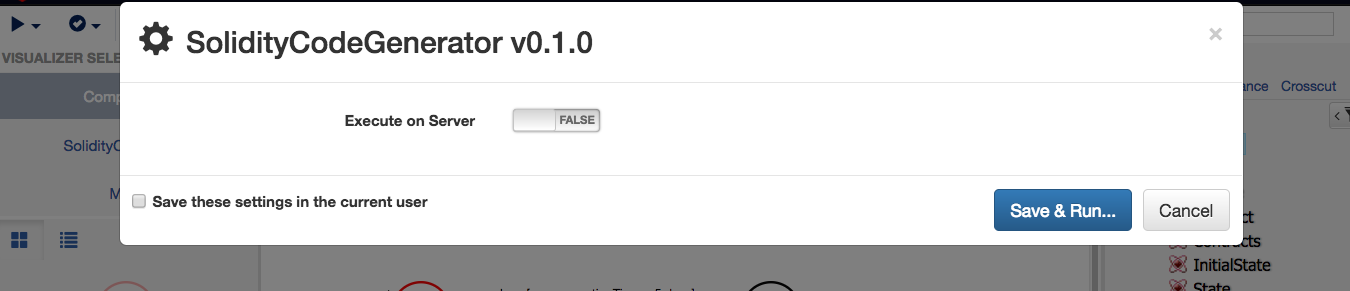}
\caption{Running the \texttt{SolidityCodeGenerator}}
\label{fig:generator}
\end{figure}

\begin{figure}
\centering
\includegraphics[width=\textwidth]{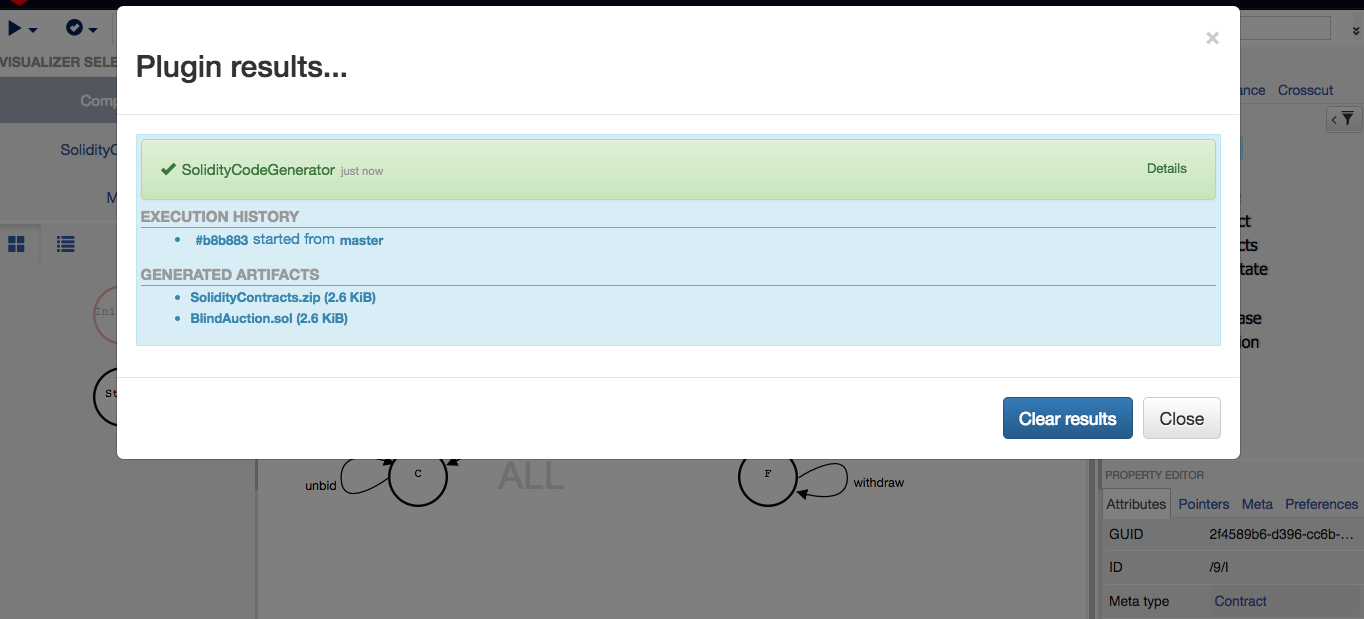}
\caption{Successful Solidity code generation}
\label{fig:correct}
\end{figure}

\begin{figure}
\centering
\includegraphics[width=\textwidth]{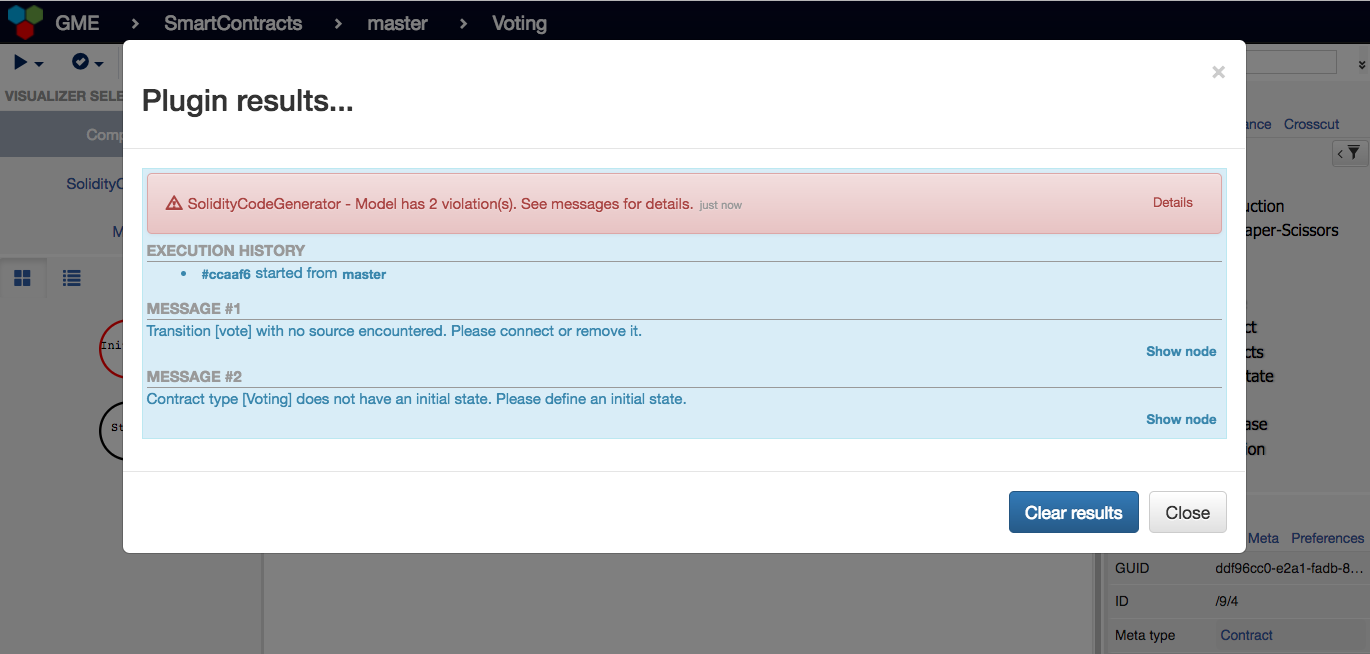}
\caption{Unsuccessful Solidity code generation due to incorrect input}
\label{fig:wrong}
\end{figure}

\begin{figure}
\centering
\includegraphics[width=\textwidth]{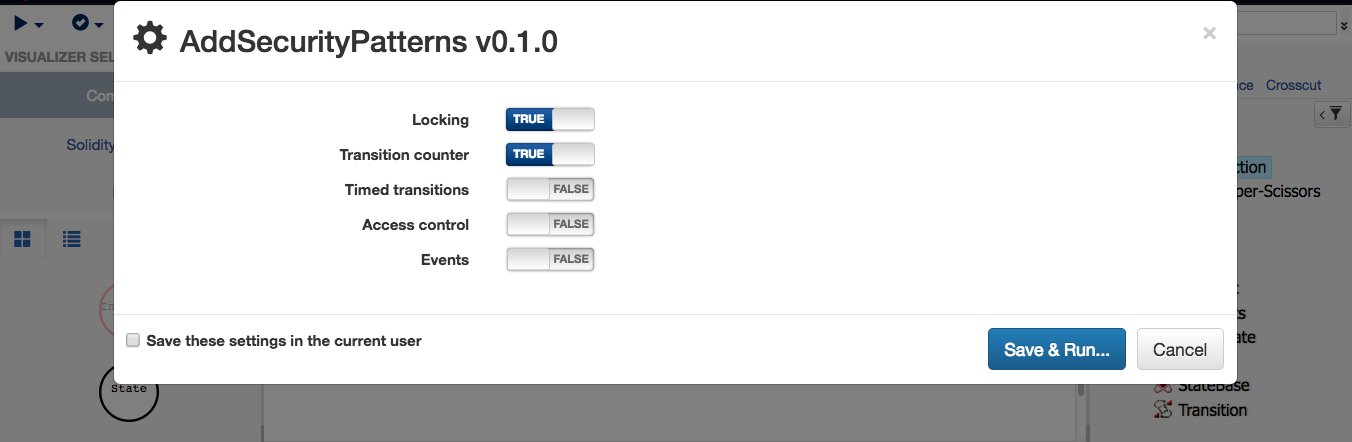}
\caption{Running the \texttt{AddSecurityPatterns}}
\label{fig:patterns}
\end{figure}
\fi

\end{document}